\documentstyle[twocolumn,floats,aps,prl,amsfonts,amssymb]{revtex}

\input{epsf}

\newcommand{\rsn}{R_{\text{SN}}}              
\newcommand{\ropt}{\rsn^{\text{opt}}}         
\newcommand{\rin}{\rsn^{\text{in}}}           
\newcommand{\mean}[1]{\langle #1\rangle}      
\newcommand{\sigopt}{\sigma_r^{\text{opt}}}   
\newcommand{\omopt}{\Omega^{\text{opt}}}      

\begin{document}
\title{Signal Selection Based on Stochastic Resonance}
\author{Hans E.\ Plesser%
        \footnote{Corresponding author, plesser@chaos.gwdg.de}
        and Theo Geisel}       
\address{Max-Planck-Institut f\"ur Str\"omungsforschung
         and Fakult\"at f\"ur Physik, Universit\"at G\"ottingen\\
         Bunsenstra\ss{}e~10, 37073~G\"ottingen, Germany}
\date{\today}
\maketitle


\begin{abstract}
  Noise aids the encoding of continuous signals into pulse sequences
  by way of stochastic resonance and endows the encoding device with a
  preferred frequency.  We study encoding by a threshold device based
  on the Ornstein-Uhlenbeck process, equivalent to the leaky
  integrate-and-fire neuron model.  Preferred frequency, optimum noise
  intensity, and optimum signal-to-noise ratio are shown to be
  linearly related to the AC amplitude of the input signal.  The DC
  component of the input tunes the device either into
  \emph{transmission} (preferred frequency nearly independent of
  signal amplitude) or \emph{selection mode} (frequency rising with
  amplitude).  We argue that this behavior may facilitate selective
  signal processing in neurons.
\end{abstract}




\section{Introduction}\label{sec:intro}

Our view of noise has shifted markedly over the past two decades:
After it had long been seen merely as a nuisance, geophysicists first
suggested in 1981 that noise may amplify the effect of weak periodic
signals on dynamic systems \cite{Benz:1981(L453)}.  This effect,
called \emph{stochastic resonance} (SR), has since been found
in a variety of experiments.  The origin of stochastic resonance
in both dynamical and non-dynamical systems is well understood today,
although theories are mostly confined to slow signals.  A recent
review of the field can be found in Ref.~\cite{Gamm:1998(223)}.

The finding that noise may aid signal detection and transmission has
spurred intense research in the neurosciences, where scientists have
long been puzzled by the seeming irregularity of neuronal activity.
The benefits of noise for signal processing in neurons have now been
demonstrated in a wide range of species: neurons transduce a signal
(or stimulus) optimally if a certain amount of ambient noise is
present 
\cite{ref3-10};
see Ref.~\cite{Wies:1998(539)} for a review.  

Recent studies by ourselves \cite{Ples:1999(7008)} and other authors
\cite{ref13-14} have revealed a further noise-induced resonance effect
in model neurons: There is also an optimum signal frequency, for which
the neuron responds with spike trains (output signals) that have a
particularly high signal-to-noise ratio (SNR).  In this letter, we
show that this resonance frequency and the optimum noise amplitude are
linearly related to the AC component of the signal impinging on a
neuron, while the DC component serves as tuning parameter.  We argue
that these relations permit the neuron to switch between a signal
transmission and a signal selection mode of operation.

The present work is based on the integrate-and-fire neuron model,
which will briefly be reviewed in Sec.~\ref{sec:model}, before the
noise-induced response properties of the neuron are presented in
Sec.~\ref{sec:response}.  Implications for neuronal signal processing
are discussed in Sec.~\ref{sec:function}.  More details are contained
in Ref.~\cite{Ples:PhD}.


\section{Model}\label{sec:model}

We focus on the spike generator of neurons for the sake of
both simplicity and generality.   The spike generator integrates the
net input current $I(t)$ impinging on the neuron like a leaky
capacitor.  For sinusoidal input superimposed with white noise, the
potential $v(t)$ across the capacitor (the membrane potential) is thus
governed by \cite{Tuck:Stoc}
\begin{equation}
  \dot{v}(t) = -v(t) + \mu + q\cos(\Omega t + \phi) + \sigma\xi(t).
  \label{eq:langevin}
\end{equation}
The input is characterized by the DC offset $\mu$, the signal
amplitude $q$, the frequency $\Omega$, and the (arbitrary) initial
phase $\phi$. The noise term with root mean square amplitude $\sigma$
and autocorrelation $\mean{\xi(t)\xi(t')}=\delta(t-t')$ subsumes the
noise arising both from cell biochemistry \cite{Manw:1999(1797)} and
background activity in the neuronal network
\cite{Main:1995(1503)}.  The neuron emits a stereotyped
voltage pulse (a spike) whenever the membrane
potential reaches a firing threshold $v(t_k)=1=\Theta$; the
potential is reset to $v(t_k^+)=v_r < 1$ immediately thereafter.  Time
and voltage are measured in their natural units, i.e., the
membrane time constant $\tau_m$ and the firing threshold $\Theta$.

The spike generator thus operates as an analog-to-digital converter
\cite{Sarp:1998(1601)}, encoding the continuous input signal $I(t)$
into a pulse train $f(t) = \sum_k \delta(t-t_k)$. Even though this
model is a gross simplification of biological neurons, it has proven most
fruitful for investigations of the nature of the neuronal code
\cite{ref20-24,Troy:1997(971)}.
For a derivation of this model from more realistic
neuron models, see
\cite{ref26-28}.

We restrict ourselves to sub-threshold signals, which would
not elicit any spikes in the absence of noise, i.e.,
$\max_{t\to\infty} v(t) = \mu+q/\sqrt{1+\Omega^2} < 1$ from
integration of Eq.~(\ref{eq:langevin}).  These appear
to be more relevant than super-threshold signals for the encoding of
periodic signals \cite{Kemp:1998(1987)}.  Note that the membrane
potential will oscillate about $\bar{v}=\mu$ after an initial
transient in the absence of noise.

We measure the performance of the neuron in coding the sinusoidal
input by the signal-to-noise ratio
\begin{equation}
     \rsn = {S(\Omega)}/{S_{\text{P}}}
  \label{eq:snr}
\end{equation}
of the output spike train at the signal frequency.  Here,
$S(\Omega)=\frac{1}{\pi T_o}|\int_0^{T_o} f(t) e^{i\Omega t} dt|^2$ is
the power spectral density of the train for a given observation time
$T_o$, while the white power spectrum of a Poissonian spike train with
equal mean interspike interval $\mean{\tau}$, i.e.,
$S_{\text{P}}=(\pi\mean{\tau})^{-1}$, is used as reference noise
level.  $T_o=200$ is employed throughout \cite{Ples:1999(7008),method}.


\section{Response Properties}\label{sec:response}

\begin{figure}
  \centerline{\epsfbox{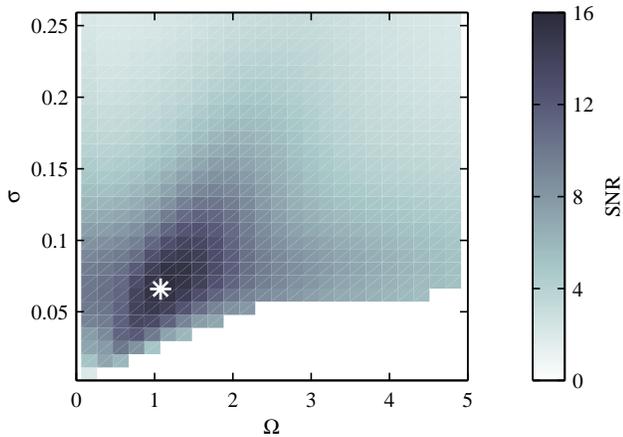}}
  \caption{Signal-to-noise ratio $\rsn$ as function of signal
    frequency $\Omega$ and input noise amplitude $\sigma$ shown
    as grayscale plot.  The asterisk marks
    $\ropt=15.7$.      No SNR could be determined for the white area
    to the bottom-right, since the
    neuron is practically quiet there.  Other parameters:
    $q=0.1$, $\mu=0.9$, $v_r=0$, and thus $\gamma=9$.}
  \label{fig:sdr}
\end{figure}

The leaky integrate-and-fire neuron responds best to sinusoidal
stimulation---i.e.\ attains a maximum signal-to-noise ratio---at a
particular combination of signal frequency $\Omega$ and input noise
amplitude $\sigma$, see Fig.~\ref{fig:sdr}.  The location of the SNR
maximum is marked by an asterisk \cite{method}. In what
follows, we shall explore the location of and the SNR value at this
maximum as a function of the remaining stimulus parameters, namely the
signal amplitude $q$, the DC offset $\mu$, and the reset potential $v_r$.

In the absence of noise, sub-threshold stimuli evoke membrane
potential oscillations about $\bar{v}=\mu$ as pointed out in
Sec.~\ref{sec:model}.  The gap between this average potential and the
threshold, $\Theta-\bar{v}=1-\mu$, needs to be bridged by the
concerted effort of signal-induced oscillations and noise.  It is
therefore plausible to scale both stimulus and noise amplitude by this
threshold distance, i.e., to define relative amplitudes,
\begin{equation}
     q_r = \frac{q}{1-\mu} \quad \text{and} \quad 
     \sigma_r = \frac{\sigma}{1-\mu}.
     \label{eq:amps}
\end{equation}
Furthermore, the reset potential $v_r$ should enter only as the ratio
of the reset distance to threshold distance, which we shall refer to
as \emph{relative reset}
\begin{equation}
    \gamma = \frac{\mu - v_r}{1-\mu}.
 \label{eq:gamma}
\end{equation}
This scaling can be established rigorously via escape noise
approximations to the dynamics of the integrate-and-fire neuron
\cite{Ples:PhD,Ples:2000(367)}.

\begin{figure}
  \centerline{\epsfbox{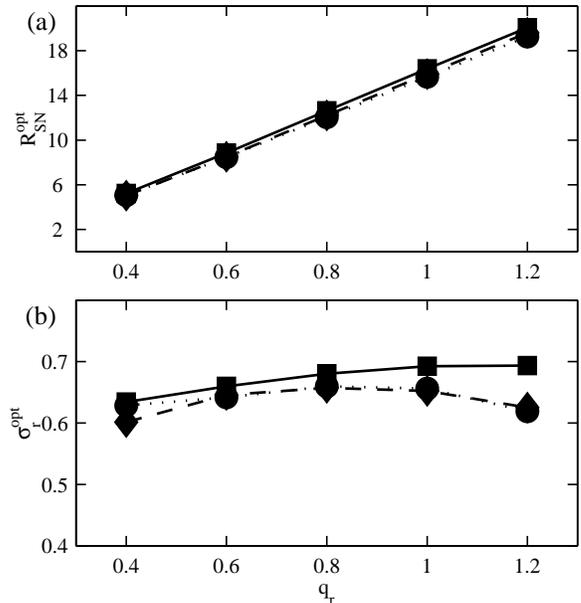}}
  \caption{(a)~Optimal
    signal-to-noise ratio and (b)~optimum noise amplitude as
    functions of the input amplitude.  Squares and solid lines mark DC
    offset $\mu=0.6$, diamonds/dashed $\mu=0.667$ and circles/dotted
    $\mu=0.889$; reset potential is $v_r=0$, yielding relative resets
    of $\gamma=1.5$, $2$, and $8$.  Lines are to guide the eye.}
  \label{fig:snrsig}
\end{figure}

We shall now turn to the relation of the optimum tuning parameters
(preferred signal frequency $\omopt$, scaled noise amplitude
$\sigopt$, and SNR $\ropt$) to the signal amplitude parameters (scaled
AC amplitude $q_r$, and relative reset $\gamma$; or DC offset
$\mu$ and reset potential $v_r$ instead of $\gamma$).
Figure~\ref{fig:snrsig}(a) indicates a perfect linear relation
between $\ropt$, the attainable SNR, and the stimulus amplitude $q_r$,
while the optimum input noise amplitude $\sigopt\approx 0.6-0.7$ is
practically independent of $q_r$, see Fig.~\ref{fig:snrsig}(b):
variations of $\sigopt$ are about one order of magnitude smaller than
the range of $q_r$ values.  Both relations are remarkably independent
of the value of the DC offset $\mu$ (indicated by symbol/line type in
Fig.~\ref{fig:snrsig}, supporting the scaling given in
Eq.~(\ref{eq:amps}).  $\ropt$ and $\sigopt$ are thus independent of the
DC component of the signal transmitted for fixed reset potential.

\begin{figure}
  \centerline{\epsfbox{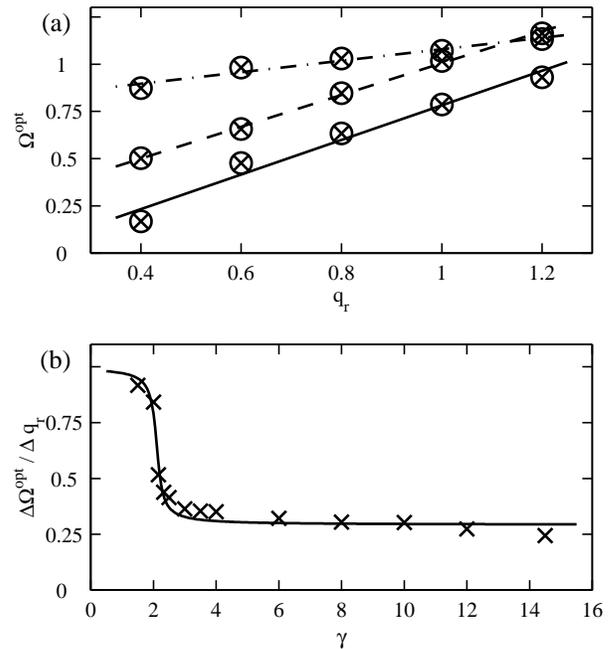}}
  \caption{%
    (a)~Optimum signal frequency as function of the input amplitude
    for relative resets $\gamma=1.5$ (solid), $\gamma=2$
    (dashed) and $\gamma=8$ (dash-dotted).  Lines are least-squares
    fits, while symbols mark different $(\mu,\, v_r)$-combinations
    yielding the same $\gamma$: circles stand for $v_r=0$ and
    crosses for $v_r=0.7$, with $\mu$ from Eq.~(\ref{eq:gamma}).
    (b)~Slope of the least-squares fits of the frequency-amplitude
    relation shown in (a) as function of the relative reset $\gamma$.
    The solid line is an empirical fit 
    $\Delta\omopt/\Delta q_r=0.65-0.23\arctan[7.0(\gamma-2.1)]$.}
 \label{fig:slope}
\end{figure}

A different behavior is observed for the optimum frequency $\omopt$ as
shown in Fig.~\ref{fig:slope}(a): For large values of $\gamma$, i.e., 
a strong positive DC offset, the optimum frequency is nearly
independent of the signal amplitude $q_r$, while small values of $\gamma$ lead
to a marked linear dependence of $\omopt$ on $q_r$: the preferred
frequency may be selected by a variation of the signal amplitude.
Figure~\ref{fig:slope}(a) also clearly indicates that the response of
the neuron depends on the DC offset $\mu$ and reset potential $v_r$
only via the relative reset $\gamma$, vindicating Eq.~(\ref{eq:gamma}):
each data point shown is a superposition of two almost perfectly
coincident points obtained from different $(\mu,\,v_r)$-combinations
yielding the same $\gamma$ (circles, crosses).  The same is found for
$\ropt$ and $\sigopt$ (not shown).  The results presented here are
thus applicable both to sensory and cortical neurons: the former are
best fit by the model for reset potentials $v_r\approx 0$
\cite{Ples:PhD}, while the latter require $v_r\approx 0.7$
\cite{Troy:1997(971)}.

Figure~\ref{fig:slope}(a) indicates that the integrate-and-fire neuron
may operate in two different modes: a \emph{transmission mode} for
large $\gamma$, which optimally encodes stimuli of a fixed preferred
frequency $\omopt\approx 1$ irrespective of their amplitude, and a
\emph{selection mode}, in which the preferred frequency may be chosen
by variation of the stimulus amplitude $q_r$.  The slope of the
frequency-amplitude curve (linear least squares fit) as a function of
the relative reset $\gamma$ is shown in
Fig.~\ref{fig:slope}(b).  There is a sharp transition between the
selection and transmission modes at $\gamma\approx 2.1$.  No slope could
be determined for $\gamma < 1.5$, since the period of the optimal
stimulus $2\pi/\omopt$ tends to the duration of the observation period
$T_o$ for small amplitudes $q_r$ in this case.

The two modes of operation arise through different firing patterns:
For large relative reset ($\gamma>2$), less than one spike is fired on
average per stimulus period, i.e., the neuron fires at most one well
phase-locked spike per period, and often skips periods in between
spikes, with a slight increase in spike number with $q_r$.
For small reset ($\gamma<2$), in contrast, the neuron has a bursting
firing pattern for small $q_r$, i.e., two or three spikes are fired in
rapid succession near the maximum of the signal in each period,
followed by silence till the next period.  As $q_r$ is increased and
the optimum frequency rises, the signal period becomes too short to
harbor more than one spike and bursting gives way to a more regular
firing pattern, with a little more than one spike per period on
average.  For intermediate reset ($\gamma\approx 2$), the neuron fires
almost regularly, with about $0.8$ spikes per signal period
independent of $q_r$.  Cold receptor neurons show all three kinds of
firing patterns (skipping, regular, bursting) depending on ambient
temperature \cite{Brau:1984(26)}; their behavior is reproduced well by
the integrate-and-fire neuron \cite{Ples:PhD}.


\section{Functional Significance}\label{sec:function}

Let us summarize the neuronal response properties and discuss them in
turn: (i)~The optimal signal-to-noise ratio scales linearly with the
input signal amplitude, and (ii)~is attained at a constant noise
amplitude, while (iii)~the preferred frequency is either independent
of (transmission mode) or linearly related to the signal amplitude
(selection mode).

Property (i)~means that the optimal SNR of the spike train emitted by the
neuron is related to the SNR of the input signal as $\ropt \sim q_r
\sim \sqrt{\rin}$ \cite{Wies:1994(2125)}, in qualitative agreement with
recent findings in a variant model neuron \cite{Burk:Sync}.  This
suggests a law of diminishing returns for the signal-to-noise ratio:
there is no point in investing valuable resources to improve $\rin$
beyond a certain level, because resulting gains in $\ropt$ would be
minimal.  Since the output of each neuron in turn is input to other
neurons, the same argument holds for raising $\ropt$.  The level of
noise observed in the brain might thus reflect an evolutionary
compromise between coding quality and resource consumption.

(ii)~The independence of the optimum input noise
amplitude $\sigopt$ from signal amplitude makes the integrate-and-fire
neuron a useful signal processing device, as no fine-tuning of the
noise to the signal is required to attain optimal performance.  The
noise level need only be adjusted relative to the DC offset, which
largely reflects homogeneous background activity.  The optimum noise
amplitude of $\sigopt\approx 0.6-0.7$ (relative to the threshold
distance $1-\mu$) is in good agreement with the observation that
coincidence detection in the auditory system of barn owls works best
for sub-threshold stimuli which raise the average membrane potential
to roughly one noise amplitude below threshold \cite{Kemp:1998(1987)}.

Property~(iii) is the central finding reported here: a model neuron as
simple as the integrate-and-fire neuron may switch between two
distinct modes of operation, a transmission and a selection mode.
Switching between the two modes is achieved by variation of the
temporally homogeneous background input to the neuron: weak background
activity activates the selection mode, a strong background the
transmission mode.  Switching between modes requires only moderate
variations of the background activity as indicated by
Fig.~\ref{fig:slope}(b).  In the former, an input signal of particular
frequency reaching the neuron through synapses far from the cell
body---and thus the spike generator---may easily be (de-)selected:
modulatory input through synapses closer to the spike generator need
only vary the amplitude $q$ of the net input current $I(t)$ to the
spike generator to tune the neuron's optimum frequency $\omopt$ either
closer to or away from the given signal frequency $\Omega$.  Selected
signals are then coded into spike trains with high signal-to-noise
ratio, i.e., trains with clear temporal structure, while deselected
signals elicit more random output.  Since pulse packets can propagate
through networks of neurons only if they are sufficiently strong and
tight \cite{Dies:1999(529)}, variation of the SNR provides a means of
gating such packets through neuronal networks.


\section{Summary}\label{sec:summary}

We have shown here that the filter properties of a threshold system
exploiting stochastic resonance in the sub-threshold regime are
linearly related to the AC amplitude of the input signal, and may be
tuned by variation of the DC signal amplitude.  Our results indicate
that such a simple device may, with the aid of noise, provide the
means to selectively transmit signals in neuronal networks. It might
thus harness noise for the benefit of neuronal computation.  Although
our study is set in the framework of neurons as the most widespread
threshold detectors in nature, the results apply more generally to any
threshold system that may be characterized as an Ornstein-Uhlenbeck
escape process.

We would like to thank A.~N.~Burkitt, G.~T.~Einevoll and W.~Gerstner
for critically reading an earlier version of the manuscript.



\begin{thebibliography}{10}

\bibitem{Benz:1981(L453)}
R. Benzi, A. Sutera, and A. Vulpiani, J. Phys. A {\bf 14},  L453  (1981).

\bibitem{Gamm:1998(223)}
L. Gammaitoni, P. H{\"a}nggi, P. Jung, and F. Marchesoni, Rev. Mod. Phys. {\bf
  70},  223  (1998).

\bibitem{ref3-10}
A. Longtin, A. Bulsara, and F. Moss, Phys. Rev. Lett. {\bf 67},  656  (1991).
J.~K. Douglass, L. Wilkens, E. Pantazelou, and F. Moss, Nature {\bf 365},  337
  (1993).
J.~E. Levin and J.~P. Miller, Nature {\bf 380},  165  (1996).
J.~J. Collins, T.~T. Imhoff, and P. Grigg, J. Neurophysiol. {\bf 76},  642
  (1996).
P. Cordo {\it et~al.}, Nature {\bf 383},  769  (1996).
F. Jaramillo and K. Wiesenfeld, Nature Neurosci. {\bf 1},  384  (1998).
E. Simonotto {\it et~al.}, Phys. Rev. Lett. {\bf 78},  1186  (1997).
R. Srebro and P. Malladi, Phys. Rev. E {\bf 59},  2566  (1999).

\bibitem{Wies:1998(539)}
K. Wiesenfeld and F. Jaramillo, Chaos {\bf 8},  539  (1998).

\bibitem{Ples:1999(7008)}
H.~E. Plesser and T. Geisel, Phys. Rev. E {\bf 59},  7008  (1999).

\bibitem{ref13-14}
F. Liu, J.~F. Wang, and W. Wang, Phys. Rev. E {\bf 59},  3453  (1999).
T. Kanamaru, T. Horita, and Y. Okabe, Phys. Lett. A {\bf 255},  23  (1999).

\bibitem{Ples:PhD}
H.~E. Plesser, Ph.D. thesis, Georg-August-Universit{\"a}t,
G{\"o}ttingen, 1999, webdoc.sub.gwdg.de/diss/1999/plesser.

\bibitem{Tuck:Stoc}
H.~C. Tuckwell, {\em Stochastic Processes in the Neurosciences} (SIAM,
  Philadelphia, 1989).

\bibitem{Manw:1999(1797)}
A. Manwani and C. Koch, Neural Comput. {\bf 11},  1797  (1999).

\bibitem{Main:1995(1503)}
Z.~F. Mainen and T.~J. Sejnowski, Science {\bf 268},  1503  (1995).

\bibitem{Sarp:1998(1601)}
R. Sarpeshkar, Neural Comput. {\bf 10},  1601  (1998).

\bibitem{ref20-24}
W. Gerstner, R. Kempter, J.~L. van Hemmen, and H. Wagner, Nature {\bf 383},
  76  (1996).
P. Mar{\v{s}}{\'a}lek, C. Koch, and J. Maunsell, 
   Proc. Natl. Acad. Sci. USA {\bf 94},  735  (1997).
G. Bugmann, C. Christodoulou, and J.~G. Taylor, Neural Comput. {\bf 9},  985
  (1997).
J. Feng, Phys. Rev. Lett. {\bf 79},  4505  (1997).
L.~F. Abbott, J.~A. Varela, K. Sen, and S.~B. Nelson, Science {\bf 275},  220
  (1997).

\bibitem{Troy:1997(971)}
T.~W. Troyer and K.~D. Miller, Neural Comput. {\bf 9},  971  (1997).

\bibitem{ref26-28}
W. Kistler, W. Gerstner, and J.~L. van Hemmen, Neural Comput. {\bf 9},  1015
  (1997).
C.~F. Stevens and A.~M. Zador,  in {\em Proceedings of the 5th Joint Symposium
  on Neural Computation} (Institute for Neural Computation, UCSD, La Jolla, CA,
  1998), pp.\ 172--177.
P. L{\'a}nsk{\'y}, Phys. Rev. E {\bf 55},  2040  (1997).

\bibitem{Kemp:1998(1987)}
R. Kempter, W. Gerstner, J.~L. van Hemmen, and H. Wagner, Neural Comput. {\bf
  10},  1987  (1998).

\bibitem{method}
$\rsn$ was determined for each $(\Omega,\,\sigma)$-combination 
shown in Fig.~\ref{fig:sdr} by numerical evaluation of
Eq.~(\ref{eq:snr}); see \cite{Ples:1999(7008),Ples:PhD} for details and
\cite{Shim:1999(3461)} for a related approach.  To obtain $\ropt$,
$\rsn$ was maximized with respect to $\Omega$ and $\sigma$ using a 
Nelder-Mead direct search algorithm \cite{Math:MATL(1998)}.

\bibitem{Ples:2000(367)}
H.~E. Plesser and W. Gerstner, Neural Comput. {\bf 12},  367  (2000).

\bibitem{Brau:1984(26)}
H.~A. Braun, K. Sch{\"a}fer, and H. Wissing, Funkt. Biol. Med. {\bf 3},  26
  (1984).

\bibitem{Wies:1994(2125)}
K. Wiesenfeld {\it et~al.}, Phys. Rev. Lett. {\bf 72},  2125  (1994).

\bibitem{Burk:Sync}
A.~N. Burkitt and G.~M. Clark, Synchronization of the neural response to noisy
  periodic input, 1999, submitted.

\bibitem{Dies:1999(529)}
M. Diesmann, M.-O. Gewaltig, and A. Aertsen, Nature {\bf 402},  529  (1999).

\bibitem{Shim:1999(3461)}
T. Shimokawa, A. Rogel, K. Pakdaman, and S. Sato, Phys. Rev. E {\bf 59},  3461
  (1999).

\bibitem{Math:MATL(1998)}
{\em MATLAB Function Reference}, The MathWorks, Inc., Natick, MA, USA, 1998.

\end{thebibliography}



%
%

\end{document}